\documentclass[aps,floatfix,superscriptaddress,notitlepage]{
revtex4}
\usepackage{amsmath,amssymb,amsfonts,graphics,graphicx,dcolumn,bm,enumerate}
\usepackage{comment,natbib,appendix}
\usepackage{multirow,color}

\usepackage{amsthm}
\usepackage{epsfig}
\usepackage{natbib}
\usepackage{hyperref}
\usepackage{graphicx}
\usepackage{epstopdf}

\newcommand{\cmp}
{\affiliation{Condensed Matter Physics Division, 
Saha Institute of Nuclear Physics, 1/AF Bidhannagar, Kolkata 700064, India.}}
\newcommand{\barasat}
{\affiliation{Barasat Government College, 10, K.N.C. Road, Barasat (N 24 
Parganas), Kolkata 700124, India.}}
 
\begin{document}

\title{Disorder induced phase transition in an opinion dynamics model: results 
in 2 and 3 dimensions}

\author{Sudip Mukherjee}%
\email[Email: ]{sudip.mukherjee@saha.ac.in} 
\barasat \cmp
\author{Arnab Chatterjee}%
\email[Email: ]{arnabchat@gmail.com} 
\cmp

\begin{abstract}
We study a model of continuous opinion dynamics with both positive and negative 
mutual interaction. The model shows a continuous phase transition between a phase with 
consensus (order) and a phase having no consensus (disorder).
The mean field version of the model was already studied. Using 
extensive numerical simulations, we study the same model in $2$ and $3$ dimensions.
The critical points of the phase transitions for various cases and the 
associated critical exponents have been estimated. The universality class of the phase 
transitions in the model is found to be same as Ising model in the respective dimensions.
\end{abstract}

\maketitle

%%%%%%%%%%%%%%%%%%%%%%%%%%%%%%%%%%
\section{Introduction}
%%%%%%%%%%%%%%%%%%%%%%%%%%%%%%%%%%%%

Social dynamics is being studied qualitatively and quantitatively, extensively at 
present~\cite{Chakrabarti2006econosocio,stauffer2006biology,CFL_RMP2009,
galam2012sociophysics,Sen:2013}. This interdisciplinary area used to be the traditional ground 
for social scientists, but has seen increasing participation of statistical physicists 
recently. Social systems are quite interesting -- they 
demonstrate rich emergent phenomena, resulting out of interaction of a large 
number of entities or agents~\cite{liggett1999stochastic}. These rich 
dynamical systems can be studied using various tools of statistical 
physics~\cite{CFL_RMP2009,Sen:2013}.

Our present study concerns the dynamics of opinions, and how consensus 
may or may not emerge out of interacting individual opinions, or, to be rather
specific, out of interaction of individuals whose opinions evolve out of 
influence of others. 
There has been a series of studies on this 
topic~\cite{lewenstein1992statistical,deffuant2000mixing,hegselmann2002opinion,
galam1982sociophysics,galam2002minority}, which have enriched our understanding 
in this regard. 
Opinions are usually modeled as discrete or continuous variables, which can 
undergo changes spontaneously or due to interaction with others, or even 
external factors. The interest lies in the study of dynamics of opinions, as 
well as the steady state properties -- a phase with a spectrum of opinions and 
another phase where the majority have similar values. In continuous opinion 
models, opinions cluster around a single value (consensus), or two (polarization) 
or can even have several values (fragmentation).

The present model is similar  to some simple models  proposed 
recently~\cite{lallouache2010opinion,
sen2011phase,biswas2011phase,biswas2011mean},
apparently inspired by the kinetic models of wealth 
exchange~\cite{Chatterjee2007,chakrabarti2013econophysics}.
A symmetry breaking transition was observed in such models: the average opinion 
is nonzero in the symmetry broken phase,  while the opinions of all individuals 
are identically zero indicating a `neutral state' in the symmetric phase. 
The parameters representing \textit{conviction} (self interaction) and 
\textit{influence} (mutual interaction) in these models were considered either  
to be uniform (scalar) or  in the generalized case different for each 
individual, i.e, as components of a vector.
There is an added feature of the randomness in the influence term which 
effectively controlled the sharpness of the phase transitions in these models.

The model we study has conviction parameter set to unity. In absence of 
interaction, opinions here remain frozen, while any interaction, however small,  
leads to a state of all individuals having extreme opinions~\cite{sen2011phase}.
Allowing for negative values in interaction (influence) takes care of the 
situation where a pair has a disagreement. This model~\cite{biswas2012disorder} 
uses the fraction of negative influences as the tuning parameter to study the 
phase transition behavior. The salient features of the phase transition in the 
mean field/infinite range version for the model is already known, including 
the universality class. There have been a few 
studies~\cite{crokidakis2012role,crokidakis2014impact,vieira2016noise,
vieira2016consequences,crokidakis2016noise,crokidakis2014phase,xiong2013competition,
chowdhury2014kinetic} extending this model to 
further realistic situations, by introduction of additional parameters. 
Although the mean field behavior of the model has been well investigated, the 
knowledge of the critical behavior of the model in finite dimensions can only ascertain its
universality class, which remains to be investigated.
In this paper, we study the model in $2$ and $3$ dimensions using extensive numerical 
simulations.

The rest of the paper is organized as follows: we introduce the model in Sec.~\ref{sec:2} and 
report our results in Sec.~\ref{sec:results}. We conclude with discussions in 
Sec.~\ref{sec:disc}.

%%%%%%%%%%%%%%%%%%%%%%%%%%%%%%%%%%%%%%%%%%%%%%%%%%%%%%%%%%%%%%%%%%%
\section{The Model}\label{sec:2}
%%%%%%%%%%%%%%%%%%%%%%%%%%%%%%%%%%%%%%%%%%%%%%%%%%%%%%%%%%%%%%%%%%%

We study a model for emergence of consensus, as introduced in 
Ref.~\cite{biswas2012disorder} (BCS model hereafter).
Let $o_i(t)$  be the opinion of an individual $i$ at time $t$.
In a system of $N$ individuals (referred to as the `society' hereafter),
opinions change out of pair-wise interactions via mutual influences/couplings 
$\mu_{ij}$ as:
\begin{eqnarray}
\label{eq:model}
 o_i(t+1) &=& o_i(t) + \mu_{ij} o_j(t).
\end{eqnarray}
One considers a similar equation for $o_j(t+1)$.
The choice of pairs $\left\{i,j\right\}$ is unrestricted, and hence
the model is originally defined on a fully connected graph, or in other words, 
of infinite range (mean field).
There is simply a pair-wise interaction and we imply no sum over the 
index $j$. Here $\mu_{ij}$ are real.
Following the above dynamics (Eq.~(\ref{eq:model})), agent $i$ updates his/her opinion by 
interacting with agent $j$ and is influenced by the mutual influence term $\mu_{ij}$.
The opinions are bounded, i.e., $-1 \le o_i(t) \le 1$. 
If the opinion value of an agent becomes higher (lower) than $+1$ ($-1$), then it is 
made equal to $+1$ ($-1$) to preserve this bound.
This bound, along with Eq.~(\ref{eq:model}) defines the dynamics of the model.
Two specific cases were studied, namely discrete and continuous $\mu$.
In the first case, the \textit{discrete} $\mu$ case, $\mu \in \{-1,+1\}$ i.e.,  
takes values $\pm 1$ only,
while in the second case, the \textit{continuous} $\mu$ case, $\mu \in 
[-1,+1]$, i.e., takes any real value between $-1$ and $+1$.
The ordering in the system is measured by the quantity $O=  |\sum_i o_i |/N$, 
the average opinion, which is the order parameter for the system. 
Changing the fraction of negative opinions $p$ one can observe a symmetry 
breaking transition between an ordered and a disordered phases -- below a 
certain value $p_c$ of the parameter $p$, the system orders (giving a non zero, 
finite value of the order parameter $O$), while a disordered phase exists above 
$p_c$ ($O=0$).
For the discrete $\mu$ case, it was shown following analytical 
calculations that the critical point is $p_c=\frac{1}{4}$ and $\beta=\frac{1}{2}$, while  
Monte Carlo simulations confirmed the results, additionally finding  $\gamma=1$ and 
$\bar{\nu}=\nu d =2$. 
$\beta$ and $\gamma$ are scaling exponents for order parameter and susceptibility respectively.
Thus $\nu=\frac{1}{2}$, if the upper critical dimension is taken as $d=4$.
However, for the continuous $\mu$ case, Monte Carlo simulations gave $p_c 
\approx 0.34$, while the critical exponents were the same as in the discrete case.

\section{Results}\label{sec:results}
In one dimension, the model shows no phase transition at non-zero value of $p$.
In our study, we investigate the same model in $2$ and $3$ dimensions.
We imagine agents to be fixed at the vertices of a hypercubic lattice of dimension $d$. During 
the dynamics, a vertex $i$ is chosen at random and one of its $2d$ neighbors ($j$) is 
randomly chosen to interact according to Eq.~(\ref{eq:model}). 
For Monte Carlo simulations in $2$ and $3$-dimensions, we realize the model on a square 
lattice and a cubic lattice respectively, and use helical boundary conditions. 
We simulate both the discrete and the continuous $\mu$ models.

We perform Monte Carlo simulations of the BCS model~\cite{biswas2012disorder} 
in 2 and 3 dimensions. The observed phase transition is quite similar to 
a thermally driven ferromagnetic-paramagnetic transition in magnetic systems. 
We compute the following quantities:
\begin{enumerate}
 \item [(a)] the average order parameter $\langle O \rangle$, $\langle \ldots 
\rangle$ means average over configurations.

\item [(b)] $V=N \left[ \langle O^2 \rangle - \langle O \rangle^2 \right]$, 
analogous to susceptibility per agent,

\item [(c)] $U= 1- \frac{\langle O^4 \rangle}{3\langle O^2 \rangle^2 }$, the 
fourth order Binder cumulant.

\end{enumerate}
The critical point is calculated from the crossing of the Binder cumulant curves 
$U(L,p)$ for different system sizes $L$. It is known that the value of the Binder 
cumulant at the critical point for a continuous phase transition is independent 
of the system size $L$, and we call this the \textit{critical Binder cumulant} $U^*$.

The order parameter $O$ behaves as $O \sim |p-p_c|^{\beta}$ and the susceptibility $V$
as $V \sim |p-p_c|^{-\gamma}$ near the critical point, i.e., for small values of $|p-p_c|$.
In order to calculate the critical exponents, we apply finite-size scaling 
(FSS) theory. We expect, for large system sizes, an asymptotic FSS behavior
of the form
\begin{eqnarray}
 O= L^{-\beta/\nu} \mathcal{F}_O (x) \left[ 1+\ldots \right]\\
 V= L^{\gamma/\nu} \mathcal{F}_V (x) \left[ 1+\ldots \right],
\end{eqnarray}
where $\cal{F}$ are scaling functions with $x= (p-p_c)L^{1/\nu}$
as the scaling variable.
The dots $[1 + \ldots]$ indicate the corrections to scaling terms.

\subsubsection{$2$ dimensions}
For 2-dimensions, we simulated the model for system sizes $N=L^2$, with 
$L=12,16,24,32,48,64,96$, with  averages over a number of configurations
ranging from $2000$ for $L=12$ to $1000$ for $L=96$.

\begin{figure}[t]
\includegraphics[width=5.9cm]{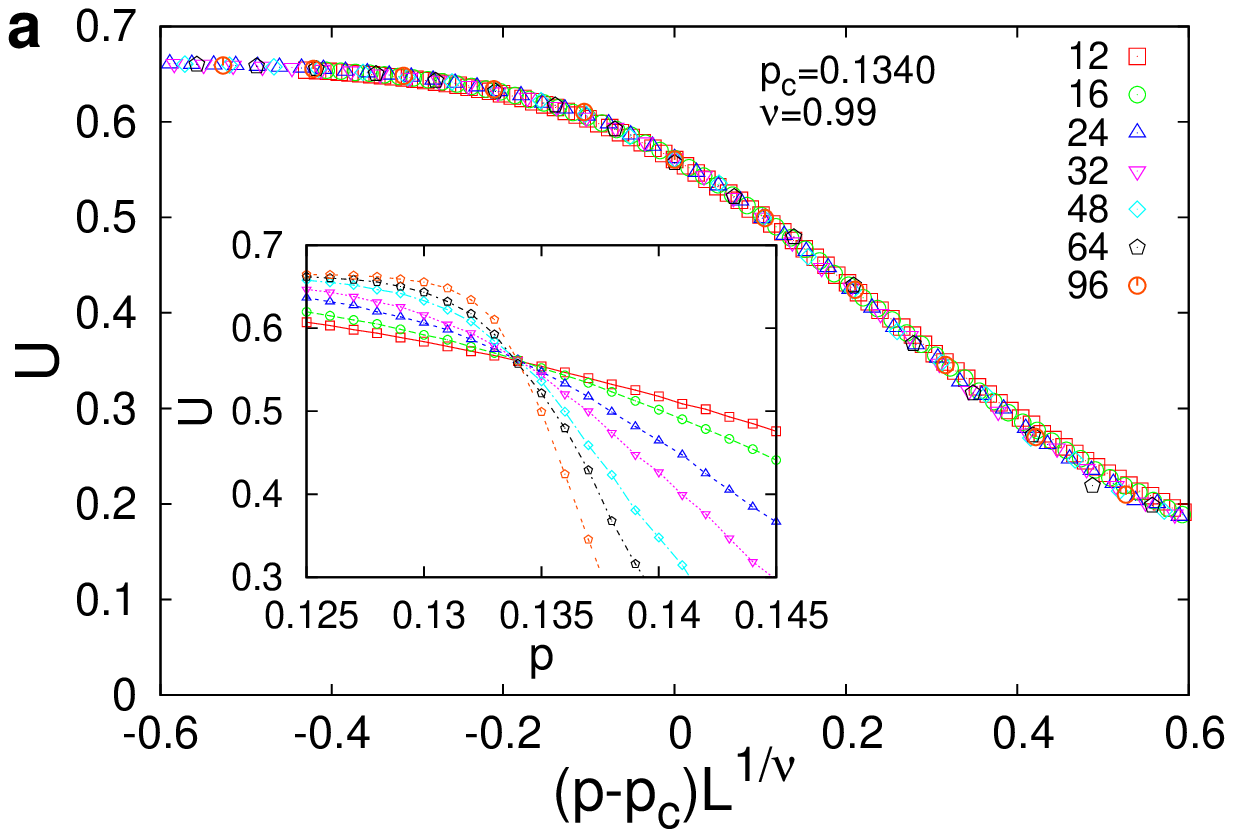}
\includegraphics[width=5.9cm]{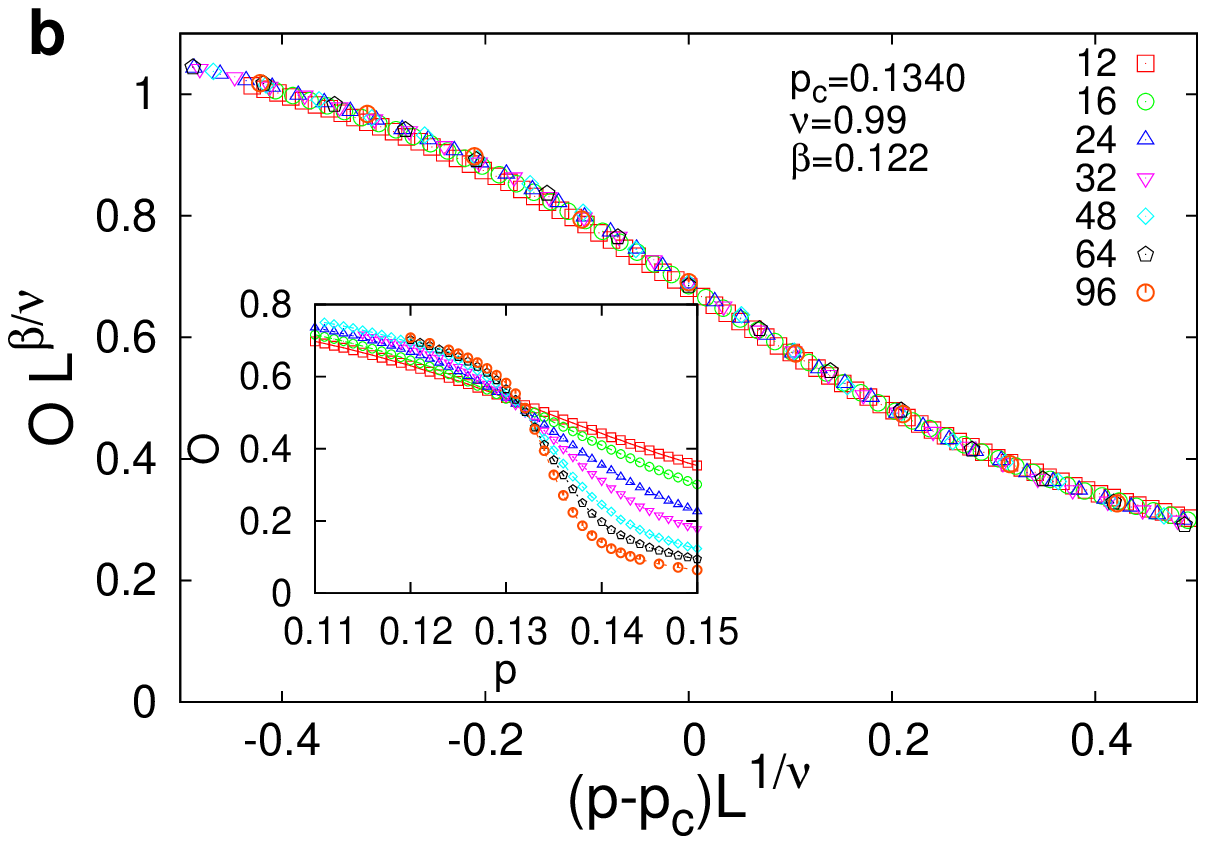}
\includegraphics[width=5.9cm]{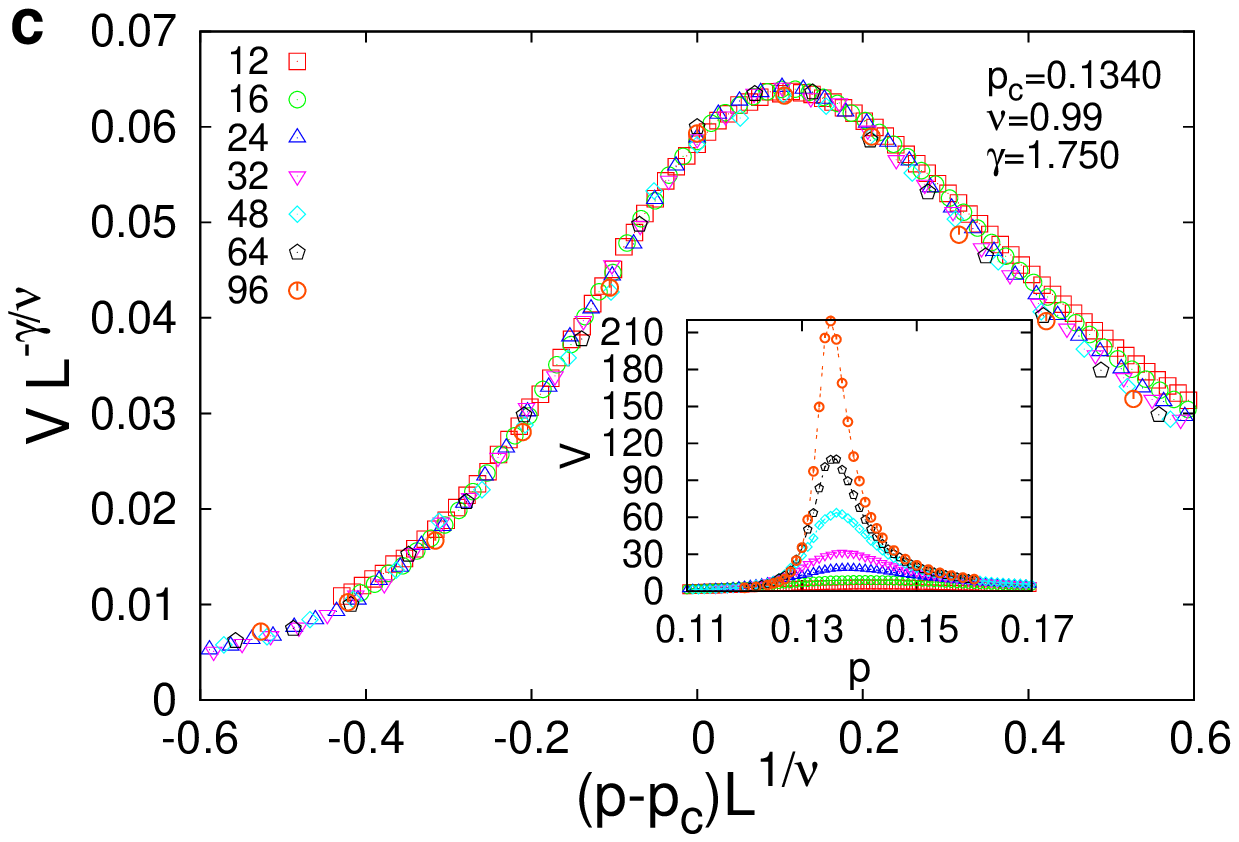}
\caption{Finite size scaling behavior for discrete $\mu$ case in $d=2$:
(a) Scaling collapse of Binder cumulant with $p_c=0.1340 \pm 
0.0001$  estimated from the crossing for different sizes $L$ (inset). 
$\nu=0.99 \pm 0.01$ is estimated from the scaling collapse. 
Critical Binder cumulant value is $U^* = 0.561 \pm 0.001$.
(b) Scaling collapse of order parameter $O$ for $\beta = 0.122 \pm 0.002$.
Inset shows unscaled data for $O$ with $p$.
(c) Scaling collapse of $V$ with $\gamma = 1.75 \pm 0.01$.
Inset shows unscaled data for $V$ with $p$.
 }
 \label{fig:2dbcs_d}
\end{figure}

For the discrete $\mu$ case, we estimated 
$p_c=0.1340 \pm 0.0001$ with critical Binder cumulant $U^* = 0.561 \pm 0.001$ 
(Fig.~\ref{fig:2dbcs_d}a).
The respective critical exponents were estimated to be $\nu=0.99 \pm 0.01$,
$\beta = 0.122 \pm 0.002$ (Fig.~\ref{fig:2dbcs_d}b) and $\gamma = 1.75 \pm 
0.01$ (Fig.~\ref{fig:2dbcs_d}c).

\begin{figure}[t]
\includegraphics[width=5.9cm]{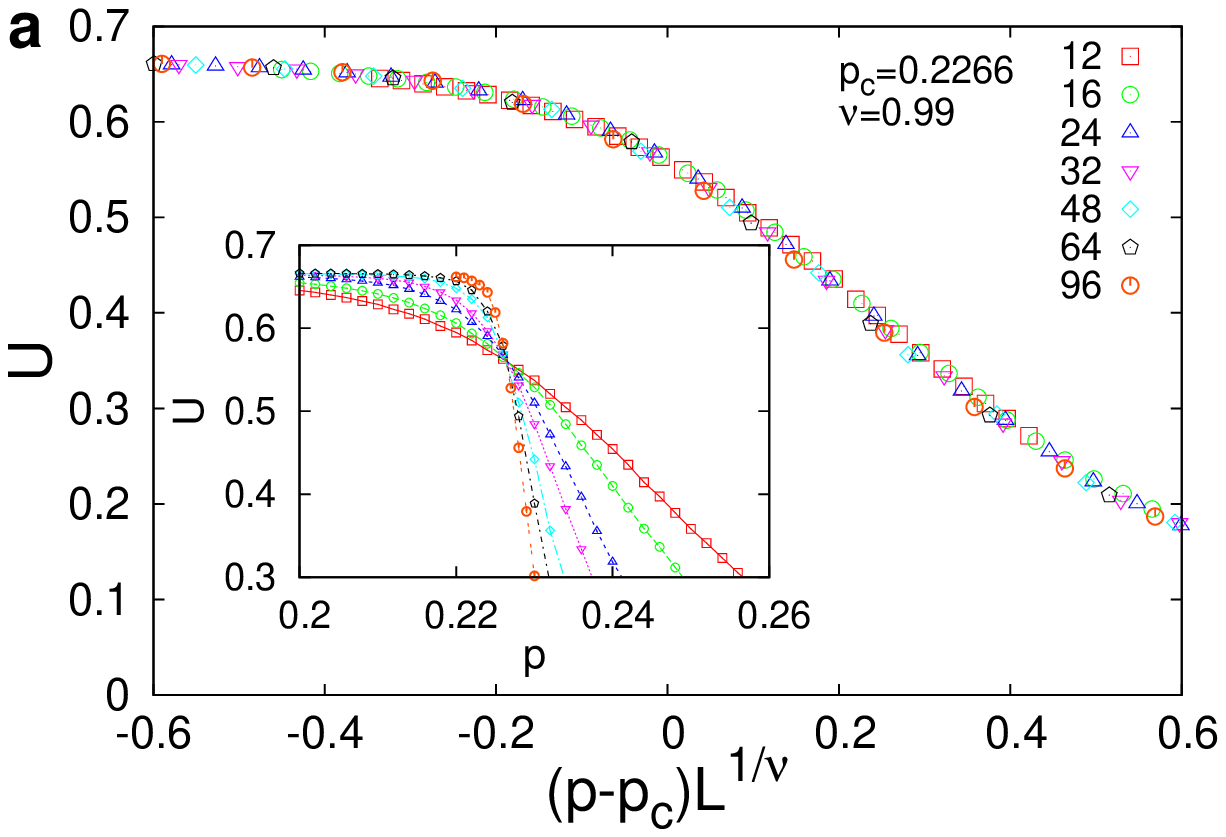}
\includegraphics[width=5.9cm]{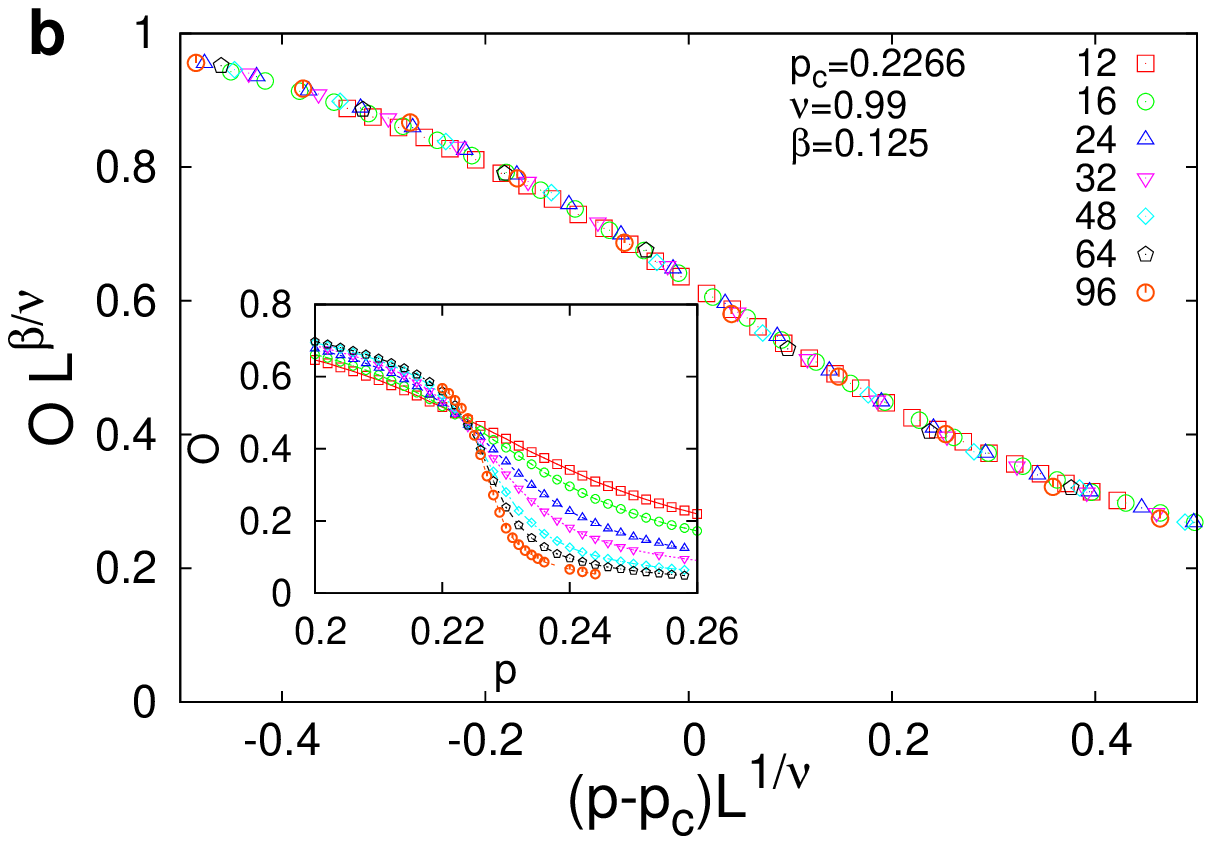}
\includegraphics[width=5.9cm]{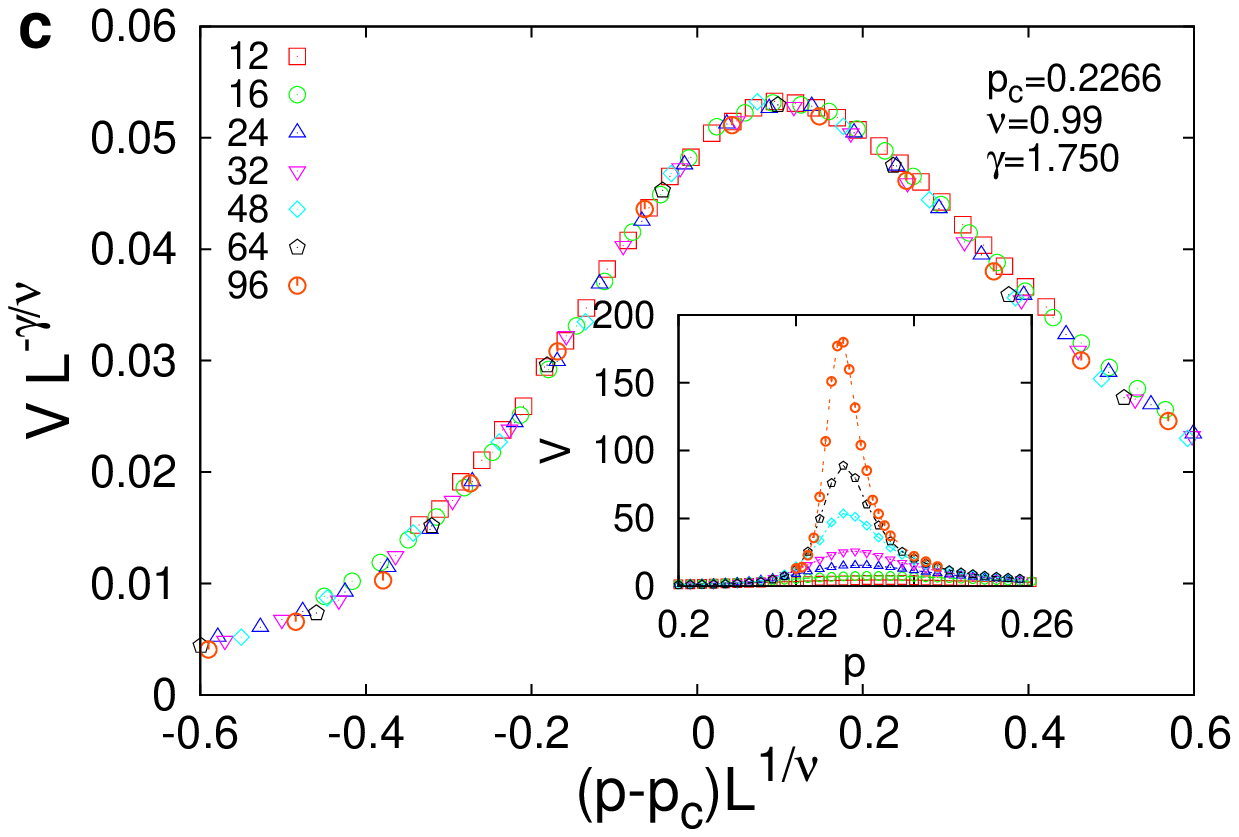}
\caption{Finite size scaling behavior for continuous $\mu$ case in $d=2$:
(a) Scaling collapse of Binder cumulant with $p_c=0.2266 \pm 
0.0001$  estimated from the crossing for different sizes $L$ (inset). 
$\nu=0.99 \pm 0.01$ is estimated from the scaling collapse. 
Critical Binder cumulant  value is $U^* = 0.559 \pm 0.001$.
(b) Scaling collapse of order parameter $O$ for $\beta = 0.125 \pm 0.001$.
Inset shows unscaled data for $O$ with $p$.
(c) Scaling collapse of $V$ with $\gamma = 1.75 \pm 0.01$.
Inset shows unscaled data for $V$ with $p$.
 }
 \label{fig:2dbcs_c}
\end{figure}

For the continuous $\mu$ case, we estimated 
$p_c=0.2266 \pm 0.0001$ with critical Binder cumulant $U^* = 0.559 \pm 0.001$ 
(Fig.~\ref{fig:2dbcs_c}a).
The respective critical exponents were estimated to be $\nu=0.99 \pm 0.01$,
$\beta = 0.125 \pm 0.001$ (Fig.~\ref{fig:2dbcs_c}b) and $\gamma = 1.75 \pm 
0.01$ (Fig.~\ref{fig:2dbcs_c}c).
%%%%%%%%%%%%%%%%%%%%%%%%%%%%%%%%%%%%%%%%%%%%%%%%%%%%%%%%%%
%%%%%%%%%%%%%%%%%%%%%%%%%%%%%%%%%%%%%%%%%%%%%%%%%%%%%%%%%%
%%%%%%%%%%%%%%%%%%%%%%%%%%%%%%%%%%%%%%%%%%%%%%%%%%%%%%%%%%
%%%%%%%%%%%%%%%%%%%%%%%%%%%%%%%%%%%%%%%%%%%%%%%%%%%%%%%%%%

\subsubsection{$3$ dimensions}
For 3-dimensions, we simulated the model for system sizes $N=L^3$, with 
$L=8,12,16,24,32$, with  averages over a number of configurations
ranging from $2000$ for $L=8$ to $1700$ for $L=32$.

\begin{figure}[h]
\includegraphics[width=5.9cm]{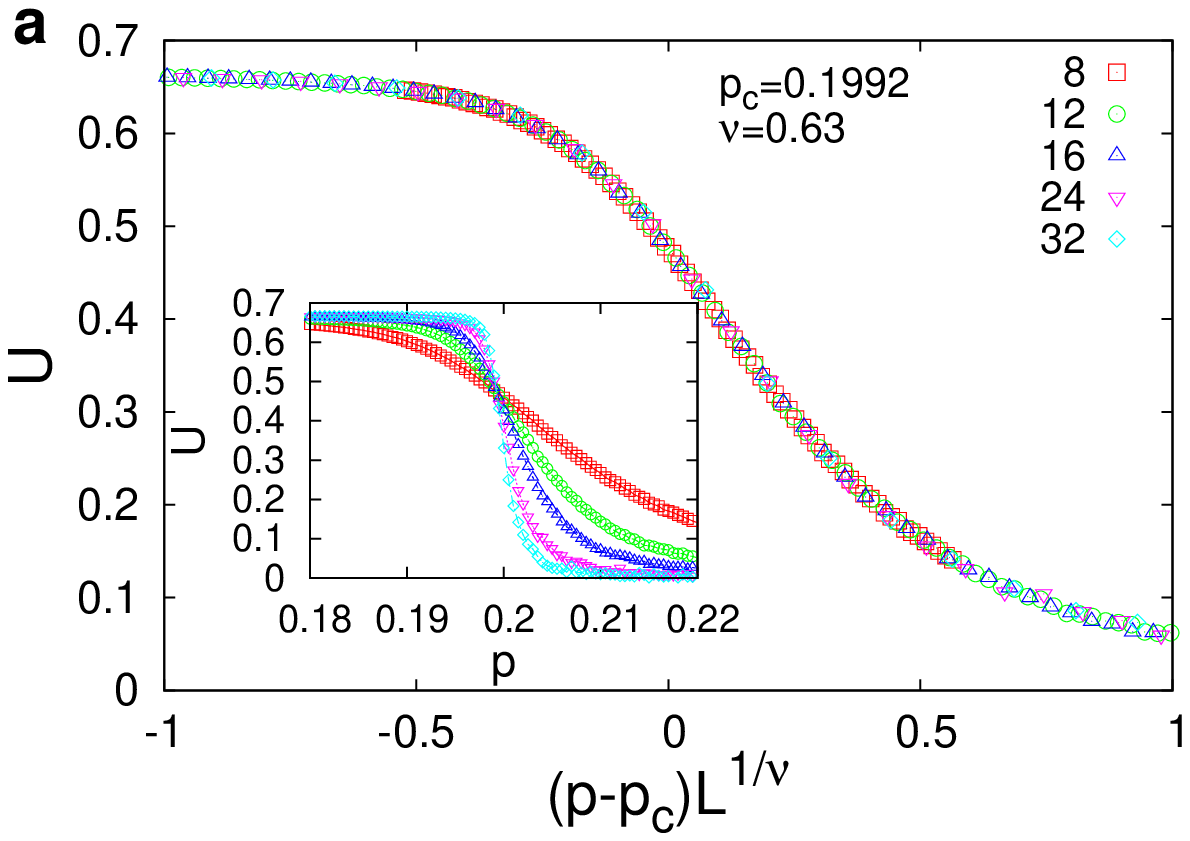}
\includegraphics[width=5.9cm]{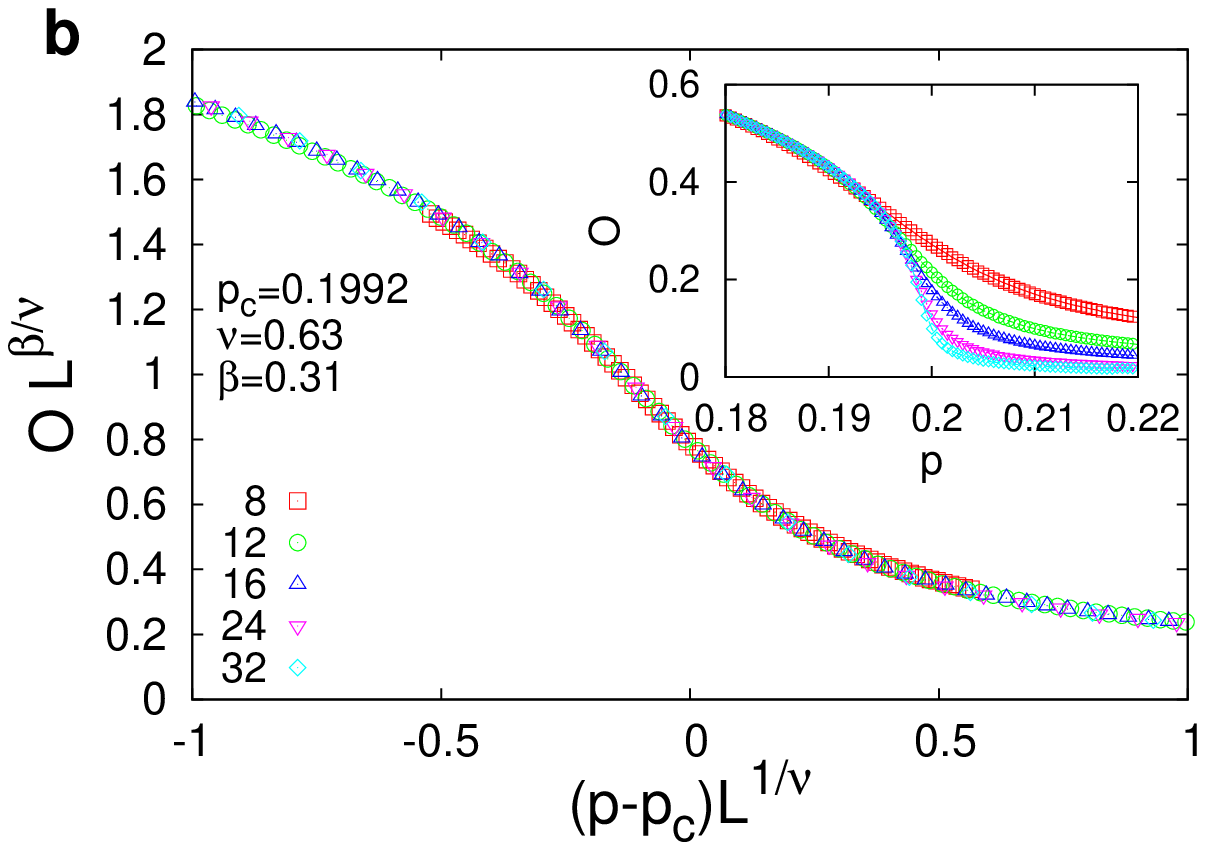}
\includegraphics[width=5.9cm]{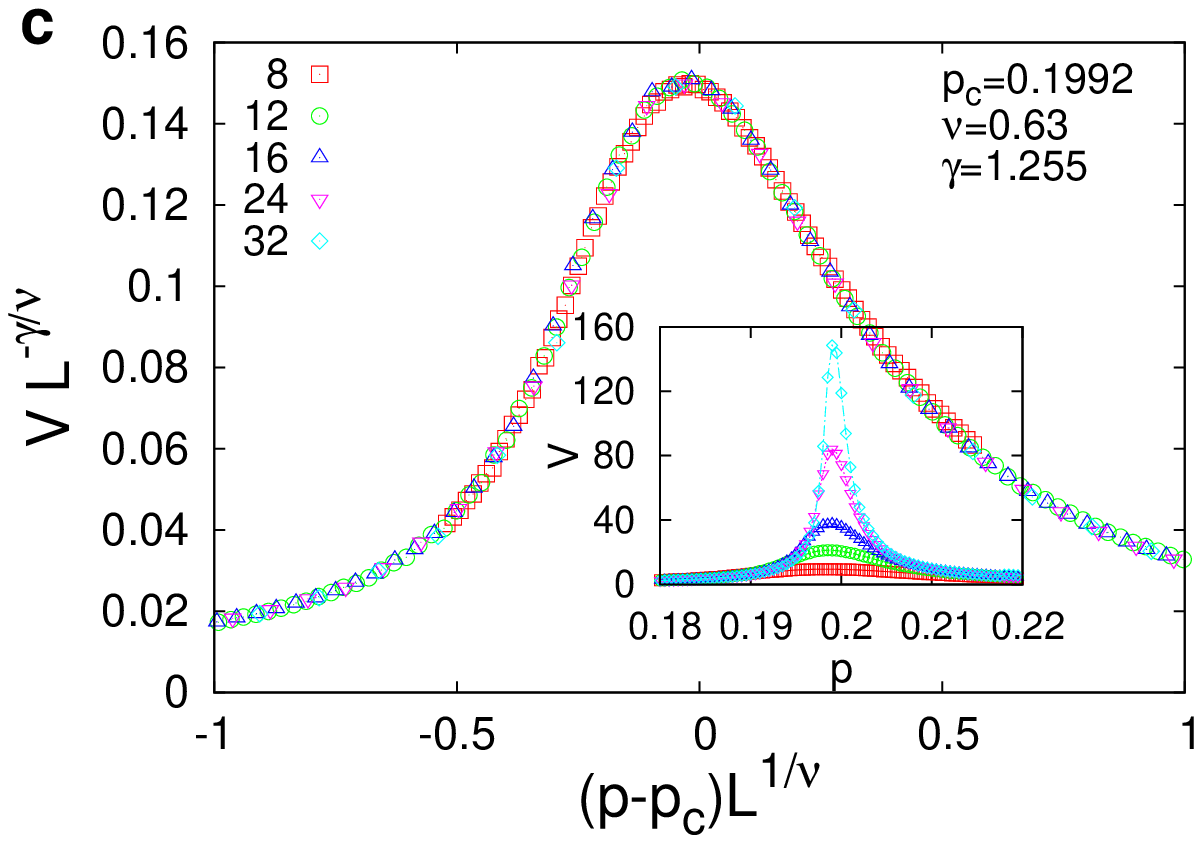}
\caption{Finite size scaling behavior for discrete $\mu$ case in $d=3$:
(a) Scaling collapse of Binder cumulant with $p_c=0.1992 \pm 
0.0002$  estimated from the crossing for different sizes $L$ (inset). 
$\nu=0.63 \pm 0.01$ is estimated from the scaling collapse. 
Critical Binder cumulant  value is $U^* = 0.476 \pm 0.004$.
(b) Scaling collapse of order parameter $O$ for $\beta = 0.310 \pm 0.002$.
Inset shows unscaled data for $O$ with $p$.
(c) Scaling collapse of $V$ with $\gamma = 1.255 \pm 0.005$.
Inset shows unscaled data for $V$ with $p$.
 }
 \label{fig:3d_d}
\end{figure}

For the discrete $\mu$ case, we estimated $p_c=0.1992 \pm 0.0002$ 
with critical Binder cumulant $U^* = 0.476 \pm 0.004$ (Fig.~\ref{fig:3d_d}a).
The respective critical exponents were estimated to be $\nu=0.63 \pm 0.01$,
$\beta = 0.310 \pm 0.002$ (Fig.~\ref{fig:3d_d}b), and $\gamma = 1.255 \pm 0.005$ 
(Fig.~\ref{fig:3d_d}c).

\begin{figure}[h]
\includegraphics[width=5.9cm]{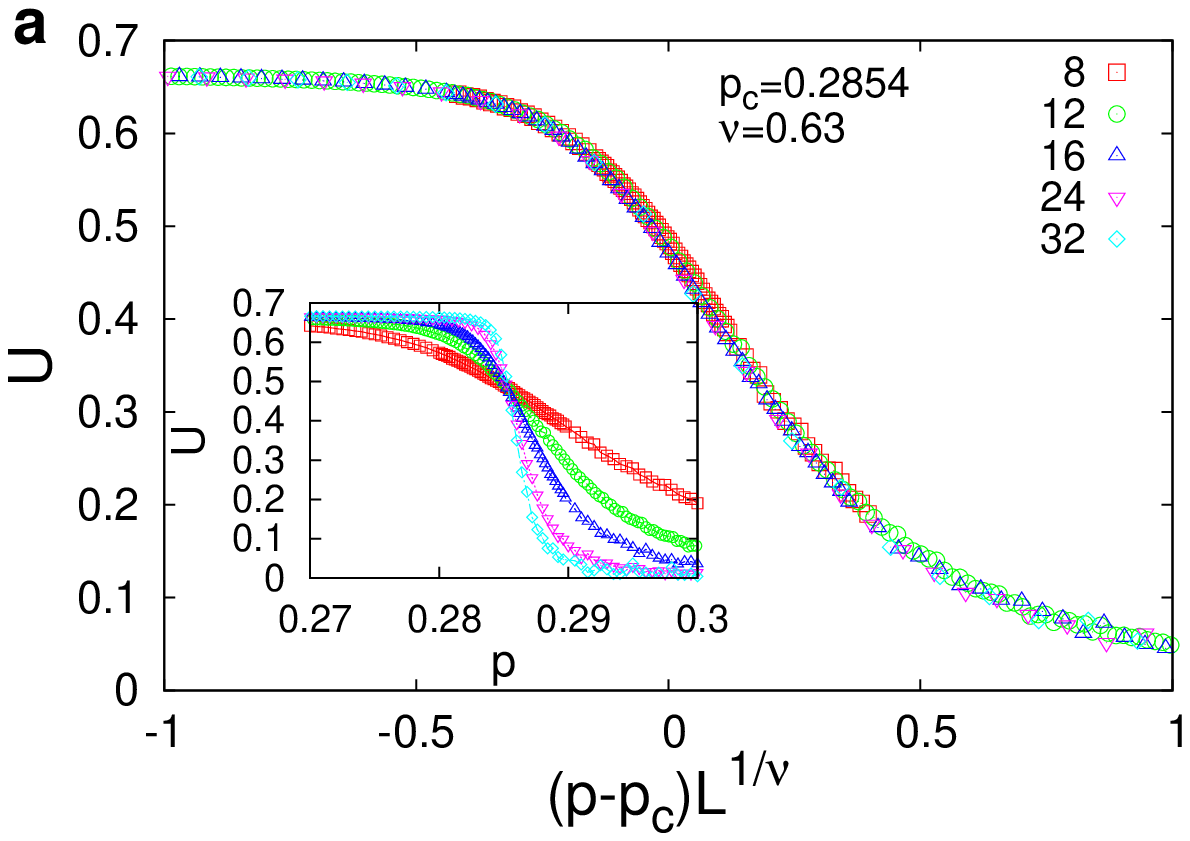}
\includegraphics[width=5.9cm]{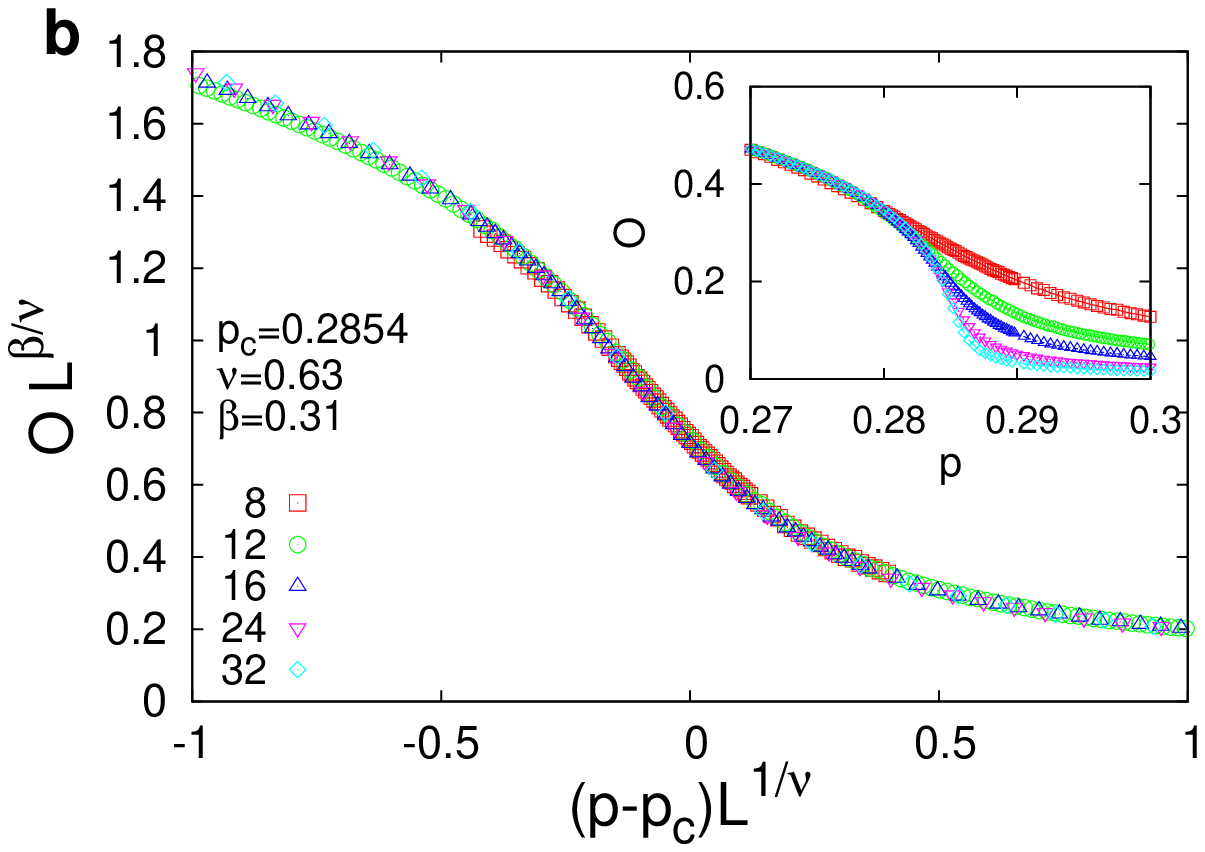}
\includegraphics[width=5.9cm]{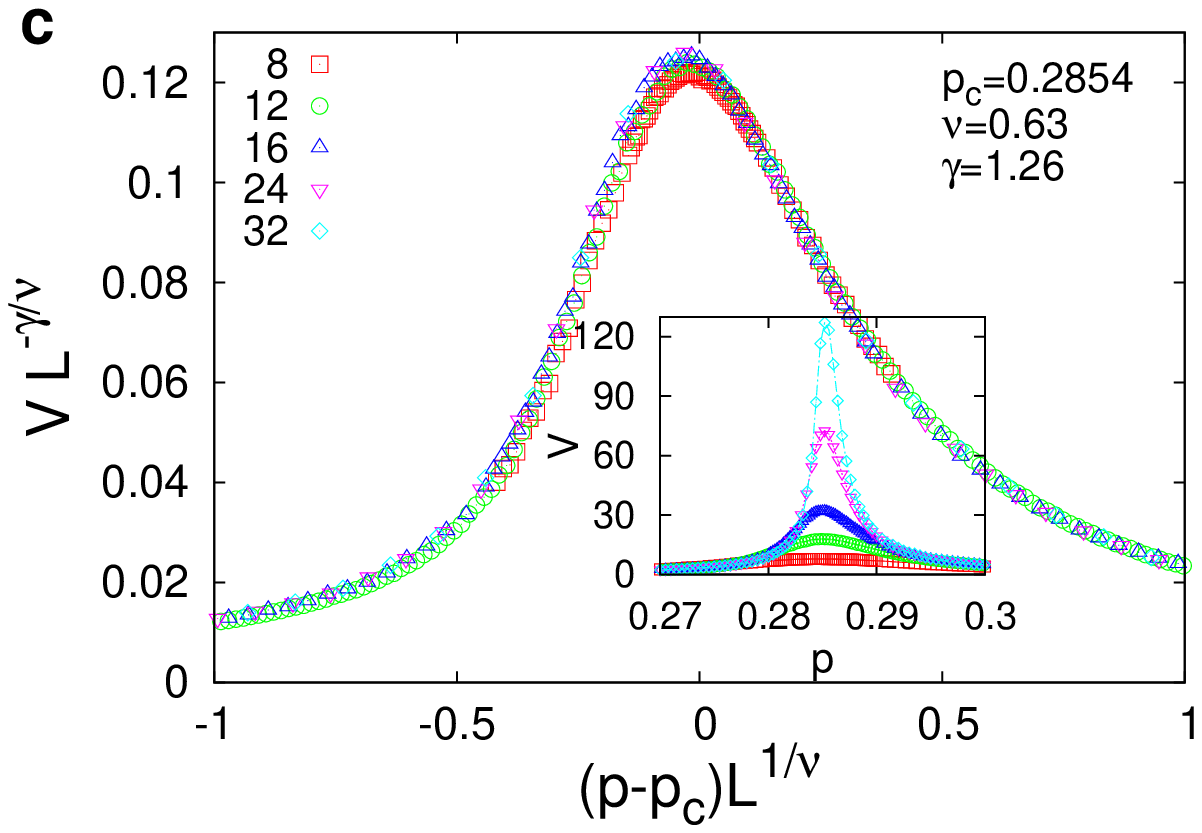}
\caption{Finite size scaling behavior for continuous $\mu$ case in $d=3$:
(a) Scaling collapse of Binder cumulant with $p_c=0.2854 \pm 
0.0001$  estimated from the crossing for different sizes $L$ (inset). 
$\nu=0.63 \pm 0.01$ is estimated from the scaling collapse. 
Critical Binder cumulant  value is $U^* = 0.485 \pm 0.002$.
(b) Scaling collapse of order parameter $O$ for $\beta = 0.310 \pm 0.002$.
Inset shows unscaled data for $O$ with $p$.
(c) Scaling collapse of $V$ with $\gamma = 1.26 \pm 0.01$.
Inset shows unscaled data for $V$ with $p$.
 }
 \label{fig:3d_c}
\end{figure}

For the continuous $\mu$ case, we estimated $p_c=0.2854 \pm 0.0001$ 
with critical Binder cumulant $U^* = 0.485 \pm 0.002$ (Fig.~\ref{fig:3d_c}a).
The respective critical exponents were estimated to be $\nu=0.63 \pm 0.01$,
$\beta = 0.310 \pm 0.002$  (Fig.~\ref{fig:3d_c}b), and $\gamma = 1.26 \pm 0.01$ 
(Fig.~\ref{fig:3d_c}c).

%%%%%%%%%%%%%%%%%%%%%%%%%%%%%%%%%%%%%%%%%%%%%%%%%%%%%%%%%
%%%%%%%%%%%%%%%%%%%%%%%%%%%%%%%%%%%%%%%%%%%%%%%%%%%%%%%%%%
%%%%%%%%%%%%%%%%%%%%%%%%%%%%%%%%%%%%%%%%%%%%%%%%%%%%%%%%%
%%%%%%%%%%%%%%%%%%%%%%%%%%%%%%%%%%%%%%%%%%%%%%%%%%%%%%%%%%

\section{Discussions}\label{sec:disc}
We have studied the BCS model of opinion dynamics in $2$ and $3$ dimensions. The model, 
originally proposed in infinite dimensions, has been extensively studied and shown to exhibit 
a continuous phase transition between a phase with order (full or partial order) and 
disorder (no consensus)~\cite{biswas2012disorder}. We used extensive Monte Carlo 
simulations to study the 
nature of the phase transition and critical behavior, estimate the associated critical 
exponents. Our findings indicate that the critical behavior of the model is same as that of 
the Ising model in the corresponding dimensions. Although the nature of randomness in 
the disorder parameter $p$ affects the critical point in both the discrete and continuous (mutual 
influence parameter) $\mu$ versions of the model, the values of the critical exponents $\nu$, 
$\beta$ and $\gamma$ are very close to the values known for the Ising model in the respective 
dimensions.

\begin{table}[h]
\caption{Comparing the critical exponents of the model studied, with Ising model in different 
dimensions. Mean field exponents for the model are taken from Ref.~\cite{biswas2012disorder}, 
while exponents of Ising model are taken from Ref.~\cite{stanley1971introduction}  ($d=2$, 
exact results) and Ref.~\cite{campostrini2002critical} ($d=3$).}
\centering
\begin{tabular}{|l|c|c|c|c|}
\hline
dimension & $p_c$ & $\nu$ &  $\beta$ & $\gamma$ \\ \hline

mean field, discrete $\mu$ & $\frac{1}{4}$ (exact); $0.250\pm 0.001$~\cite{biswas2012disorder}  
& $\nu d=2.00\pm0.01$ & $\frac{1}{2}$ (exact); $0.50 \pm 0.01$~\cite{biswas2012disorder} & 
$1.00\pm0.05$~\cite{biswas2012disorder}  \\ \hline

mean field, continuous $\mu$  & $0.3404\pm0.0002$~\cite{biswas2012disorder} & 
$\nu d=2.00\pm0.01$~\cite{biswas2012disorder}& 
$0.50 \pm 0.01$~\cite{biswas2012disorder}  &  $1.00\pm0.05$~\cite{biswas2012disorder}\\ \hline

$d=2$, discrete $\mu$ & $0.1340\pm0.0001$ & $0.99\pm0.01$ & $0.122 \pm 0.002$ & 
$1.75\pm0.01$  \\ \hline 

$d=2$, continuous $\mu$ & $0.2266\pm0.0001$ & $0.99\pm0.01$ & 
 $0.125 \pm 0.001$ & $1.75\pm0.01$  \\ \hline

$d=3$, discrete $\mu$ & $0.1992 \pm 0.0002$ & $0.63\pm0.01$  & $0.310 \pm 0.002$ & 
$1.255\pm0.005$ \\ \hline

$d=3$, continuous $\mu$ & $0.2854\pm0.0001$ & $0.63\pm0.01$ & $0.310 \pm 0.002$ &
$1.26\pm0.01$ \\ \hline \hline 

mean field Ising & & $\nu=\frac{1}{2}; d=4$  (exact)~\cite{stanley1971introduction}  & 
$\frac{1}{2}$  (exact)~\cite{stanley1971introduction} & $1$  
(exact)~\cite{stanley1971introduction}\\ \hline 

$d=2$ Ising & & $1$  (exact)~\cite{stanley1971introduction}  & 
$\frac{1}{8}$  (exact)~\cite{stanley1971introduction} & 
$\frac{7}{4}$ (exact)~\cite{stanley1971introduction}\\ 
\hline 

$d=3$ Ising & & $0.63012$~\cite{campostrini2002critical} & 
$0.32653$~\cite{campostrini2002critical} & $1.2373$~\cite{campostrini2002critical}\\ 
\hline

\end{tabular}
\label{tab:exp}
\end{table}

In Table~\ref{tab:exp}, we list the critical points and compare the critical exponents of the 
BCS model with those in the Ising model in $d=2,3$ as well as the mean field case.
However, one cannot comment on the upper critical dimension in such a model unless 
the set of critical exponents for $d \ge 4$ are computed. For example, the upper critical 
dimension of the majority voter model was found to be $6$~\cite{yang2008existence}, although 
critical exponents for that model in $2$ and $3$ dimensions reasonably matched Ising model 
exponents.
It is also important to note that the model studied is not defined by a Hamiltonian, but just 
by the microscopic dynamical rules. The opinions are not necessarily discrete, yet the 
dynamics is sufficient to demonstrate a critical behavior similar to one of the most studied 
models of statistical physics.
It may be noted that this is perhaps the first indication that a kinetic exchange type 
model can lead directly (from kinetic theory) to the celebrated cooperative (Ising) model 
universality class. This issue will  be addressed more generally in a 
forthcoming paper~\cite{MCCpap}.

\begin{acknowledgements}
The authors thank S. Biswas, B.~K. Chakrabarti and P. Sen for stimulating 
discussions and critical reading of the manuscript.
\end{acknowledgements}

%\bibliographystyle{unsrt}
%\bibliography{refbcs.bib}

\end{document}